\documentclass[prb,twocolumn]{revtex4-1}
\usepackage{epsfig}
\usepackage{amssymb}
\usepackage{amsmath}
\usepackage{graphicx}
\usepackage{amsfonts}

\begin{document}
\title{Role of Anion Ordering in the Coexistence of Spin-Density-Wave and Superconductivity in (TMTSF)$_2$ClO$_4$}
\author{Ya.\,A.~Gerasimenko$^a$, V.\,A.~Prudkoglyad$^a$, A.\,V.~Kornilov$^a$,
S.\,V.~Sanduleanu$^{a,b}$, J.\,S.~Qualls$^c$, V.\,M.~Pudalov$^{a,b}$}
\affiliation{$^a$P.\,N.~Lebedev Physical Institute of the RAS,  Moscow, 119991, Russia\\
$^b$Moscow Institute of Physics and Technology,  Dolgoprudny, 141700, Russia\\
$^c$Sonoma State University,  Rohnert Park, CA 94928, USA}

\date{\today}

\begin{abstract}
Using various transport and magnetotransport probes we study the coexistence of spin-density wave and superconductor states in (TMTSF)$_2$ClO$_4$ at various degrees of ClO$_4$ anions ordering. In the two-phase complex state when both
superconductivity and spin-density wave are observed in transport, we find prehistory effects, enhancement of the
superconducting critical field, and strong spatial anisotropy of the superconducting state. These features are inconsistent
with the conventional model of structural inhomogeneities produced by anion ordering transition.
We reveal instead that superconductor and spin-density wave regions overlap on the temperature -- dimerization gap $V$
phase diagram, where $V$ is varied by anion ordering.  The effect of anion ordering on (TMTSF)$_2$ClO$_4$ properties is thus analogous to that of pressure on (TMTSF)$_2$X (X=PF$_6$ or AsF$_6$), thereby unifying general picture of the coexistence of superconductivity and spin-density wave in these compounds.
\end{abstract}

\maketitle
\section{Introduction}

\begin{figure}[h]
\includegraphics[width=0.47\textwidth]{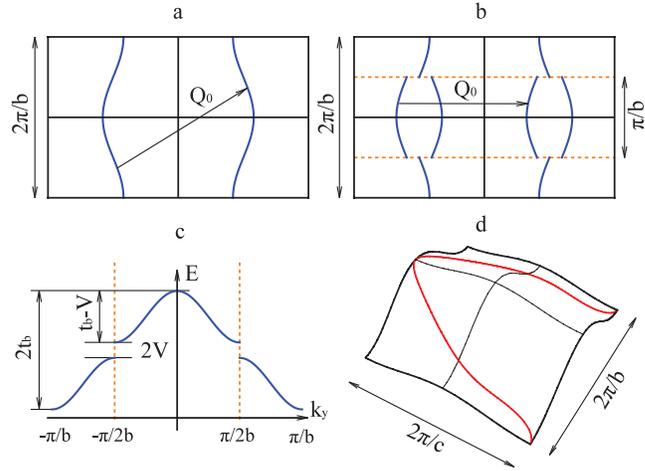}
\caption{(a) Fermi surface of PF6 or ClO4 in the absence of anion ordering. The wave vector $Q_0$ nests opposing
sheets of the Fermi surface. (b) Fermi surface of ClO4 in presence of anion ordering. Dashed line indicates the new Brillouin
zone boundary due to doubled periodicity along $\mathbf{b}$ axis. Now there are four sheets, which can be nested with $Q_0$
only when the splitting $V$ is small. (c) Sketch of the $k_y$ bandwidth measured by AMRO with or without anion ordering.
(d) One sheet of the Fermi surface with the orbit (red line), that corresponds to the magnetic field aligned in the direction
of zeroth order AMRO peak. The direction depends on $k_y$ bandwidth.}
\label{fig1}
\end{figure}

For many unconventional superconductors including  recently discovered Fe-based pnictides  \cite{reviewsFeAs} the $P-T$ phase
diagram contains overlapping  regions of the superconductor (SC) and antiferromagnetic insulator phase  (here $P$ is either
external pressure, or internal ``chemical'' pressure, induced by  chemical substitution or doping). The coexistence of the two
phases causes unusual superconducting properties in the overlap region.
Of a particular interest  are quasi one-dimensional (Q1D) organic compounds of the (TMTSF)$_2$X-family \cite{Lebed08}.
The (TMTSF)$_2$PF$_6$ (hereafter PF6) was found to demonstrate rich  physics near the spin-density wave (SDW) endpoint
\cite{Lebed08}. In the vicinity of the SDW transition the PF6 bulk homogeneous state is spontaneously split
into SDW and metal/SC areas \cite{Vuletic02,Kornilov04,Lee05} with substantially enhanced superconducting critical field
\cite{Lee02} and of strongly anisotropic  shape \cite{Kang10}.
The unusual SC behavior is believed to be due to curious properties of the SDW phase
\cite{Podolsky04,GorkovGrigoriev05,Zhang06}.
The latter is governed by nesting of the Fermi surface (FS), which easily occurs in Q1D compounds with two slightly corrugated
FS sheets (cf. Fig~\ref{fig1}a). Application of pressure increases the dimensionality of electron dispersion (via increasing
the $t_b^\prime=t_b^2/t_a$ transfer integral), which spoils nesting and progressively shifts formation of SDW to lower
temperatures.

In contrast, for apparently similar (TMTSF)$_2$ClO$_4$ (hereafter ClO4) compound, the  SC/SDW coexistence is completely
different.  ClO$_4$ anions lack inversion symmetry,  possess a dipole momentum, and experience ordering transition at
$T_{AO}=24$\,K. The transition doubles the lattice along $\mathbf{b}$ axis and splits the Fermi surface in four open sheets
(see Fig.~\ref{fig1}b). The latter spoils nesting and suppresses SDW phase giving rise to the onset of superconductor state
\cite{Lebed08}.
Due to finite time kinetics \cite{Pouget90} of the anion ordering transition, the degree of anion ordering can be varied by
adjusting the cooling rate in the vicinity of $T_{AO}$. This novel parameter enables to reveal a rich phase diagram, where SDW
phase is favored for strong disorders, SC and SDW coexist  at intermediate and SC sets in for weak disorders \cite{Schwenk84}.
Earlier measurements with ClO4 did not reveal any features similar to those for  PF6 and suggested that the SC/SDW coexistence
is due to superconducting anion-ordered inclusions in the SDW insulating disordered background \cite{Schwenk84}. This was
further supported by extensive X-ray measurements, where anion-ordered inclusions were indeed observed \cite{Pouget90}.

It was shown later, that the lower the SDW onset temperature, $T_{SDW}$, is, the stronger it is affected by magnetic
field $H||\mathbf{c}$ \cite{Matsunaga99, Matsunaga00, Qualls00}, consistent with the conventional mean-field description of
SDW \cite{Montambaux88}. This, however, implies that some parameter, that spoils FS nesting, increases when $T_{SDW}$
decreases. The latter occurs with lowering disorder, when only the FS splitting $V$ is expected to change. Indeed, $V$ plays
the role similar to $t_b^\prime$ in PF6. For the strongest disorder the Fermi surface for ClO4 is similar to that for PF6 and can,
thus, be nested (cf. Fig.~\ref{fig1}a). Gradual ordering of anions splits the FS, and $V$ represents the deviation from
perfect nesting (cf. Fig.~\ref{fig1}b). Imperfect nesting decreases $T_{SDW}$, until, at some disorder, the deviation reaches a critical value and SDW vanishes.

This  simple picture is nicely supported by the mean-field calculations of the $T_{SDW}(V)$ phase diagram
\cite{ZanchiBjelis01,SenguptaDupuis01}.
The similarity of SDW phase diagram to that of PF6 suggests, that the remarkable features of
 the SC and SDW coexistence should be observed also in ClO4, in contrast to a previously suggested picture of
 structurally inhomogeneous granular SC state \cite{Schwenk84}. Therefore,
to resolve these conflicting views, it is necessary to study in detail the SC and SDW phase diagram as a function of $V$.
Since $V$ is expected to be spatially inhomogeneous  due to structural disorder \cite{Pouget90}, it cannot be characterized
by the residual resistivity or cooling rate at $T_{AO}$ and should be measured directly.

In this paper we use several transport probes to characterize in real and momentum space the SDW and metal/SC coexistence for
variable disorder. We use angular magnetoresistance oscillations for field rotated in the $\mathbf{a}-\mathbf{c}$ plane
to probe the FS and trace changes in $V$. We indeed observe that $V$ is reduced as $T_{SDW}$ rises with increasing disorder,
in accord with the theoretical phase diagram \cite{ZanchiBjelis01,SenguptaDupuis01}. We find that the SC phase is strongly
spatially anisotropic: while transport along $\mathbf{c}$ axis manifests survival of the SC state, the $\mathbf{a}$ axis
conduction turns out ``insulating''  with increasing disorder.
For strong disorders we observed  hysteresis in  $R(T)$ behavior  and  enhancement of superconducting critical field, similar to
that observed in PF6. These results clearly demonstrate that the metal/SC and SDW regions do overlap on the $T-V$ phase
diagram and their coexistence is driven by the changes of splitting $V$ varied with the degree of anion ordering, rather than
by structural inhomogeneities produced at anion-ordering transition.

\section{Experimental}
The single crystals of typical dimensions of $1\times 0.15\times 0.05$\,mm were synthesized using a conventional
electrochemical technique. Eight $10\mu$m annealed Pt wires were glued with conducting graphite paint on two opposite crystal
faces normal to $\mathbf{c}$ axis. $R_{xx}$ and $R_{zz}$ resistances were measured along $\mathbf{a}$ and $\mathbf{c}$ axes
correspondingly using the conventional 4-wire AC technique. Samples were mounted on the $^3$He  one-axis rotator and cooled
down to 40\,K at the rate of 0.3K/min to avoid cracks.
Initially, to achieve the strongest disorder (\emph{d1}) a heat pulse was applied to the continuously cooled $^3$He bath at
1.3\,K. This resulted in the bath heating well above $T_{AO}$ and consequent rapid cooling at the rate of $\approx$100\,K/min
in the vicinity of $T_{AO}$. All the subsequent weaker disorders were obtained by annealing a sample at temperatures
lower than $T_{AO}$ \cite{Schwenk84}. The annealing procedure consisted in warming the sample to a certain temperature at the
rate of 0.2K/min and then cooling it back at the same rate. The highest annealing temperature corresponding to disorder
\emph{d5} was 22\,K. To achieve the anion-ordered state a sample was cooled from 40\,K to 20\,K at the rate of 10\,mK/min.

Samples were aligned in magnetic field by virtue of the upper critical field, $H_{c2}$, anisotropy; in the
$\mathbf{a}-\mathbf{c}$ crystal plane $H_{c2}$ is the highest for $ H||\mathbf{a}$ and the lowest for $H||\mathbf{c}$.
Therefore, $\mathbf{a}$ direction was determined using the resistance drop due to the onset of superconductivity
on the rotation curve at a field value slightly lower than $H_{c2}||\mathbf{a}$, and $\mathbf{c}$ direction was set 90 degrees
away.

\section{Results and discussion}
\begin{figure}[t]
\includegraphics[width=0.47\textwidth]{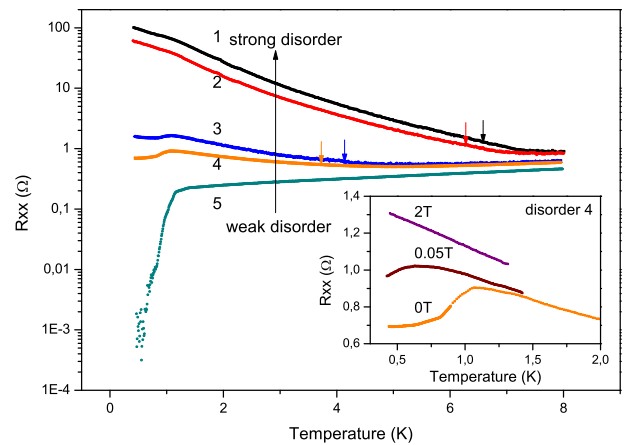}
\caption{Temperature dependences of intralayer resistance $R_{xx}$ for 5 different degrees of anion ordering. The
colors are specific to each disorder on all figures hereafter.  Arrows show the transition temperature $T_{SDW}$ defined as a
peak in the $\mathrm{d}\ln R_{xx}/\mathrm{d}(1/T)$ derivative. Inset shows the $R_{xx}(T)$ dependences in different magnetic
fields $H||\mathbf{c}$ for disorder \emph{d4}.}
\label{fig2}
\end{figure}
Figure~\ref{fig2} shows the  intralayer
$R_{xx}$ temperature dependence for various disorders. For strong disorders the sample undergoes SDW transition ($T_{SDW}=6.6$\,K
for \emph{d1}), followed by insulating temperature dependence of $R_{xx}$ for lower temperatures. Upon decreasing disorder,
SDW is suppressed, as seen from $T_{SDW}$ decrease. For disorders \emph{d3,d4} transition becomes even more smeared, however
the low-temperature behavior remains insulating and indicates the presence of SDW.

For strong disorders \emph{d1,d2},  the $R_{xx}(T)$ behavior is insulating  for $T<T_{SDW}$, although there is a kink at
$\approx 1.1$\,K which is followed by a slower growth of resistance. For weaker disorders \emph{d3, d4} the kink is followed
by resistance drop. $R_{xx}$ behavior below $T<1.1$\,K changes from yet insulating (\emph{d3}) to already metallic
(\emph{d4}), however resistance stays finite down to the base temperature 0.4\,K. Application of low magnetic field suppresses
this feature and restores insulating behavior at low temperatures, as shown in the inset to Fig.~\ref{fig2}. This indicates
that the kink and the drop are associated with the presence of superconducting inclusions within the layers in the background
SDW phase.

Finally, for the weakest disorder \emph{d5} insulating behavior is absent and superconducting transition is observed with
intralayer resistance falling to zero. The overall behavior of $R_{xx}$ is quite similar to the one observed previously by
Schwenk et al.\cite{Schwenk84}, where the authors came up with a model of a mixture of superconducting anion-ordered
inclusions in the SDW background phase. The increase in the number and size of the inclusions with decreasing disorder
(annealing the sample) leads to the percolating SC paths, the behavior similar to granular SC. The kink and the drop are then
associated with the SC transition inside the inclusions, and the metallic temperature dependence that follows the drop
indicates the developing percolating transition between the inclusions within the layers.

\begin{figure}[t]
\includegraphics[width=0.47\textwidth]{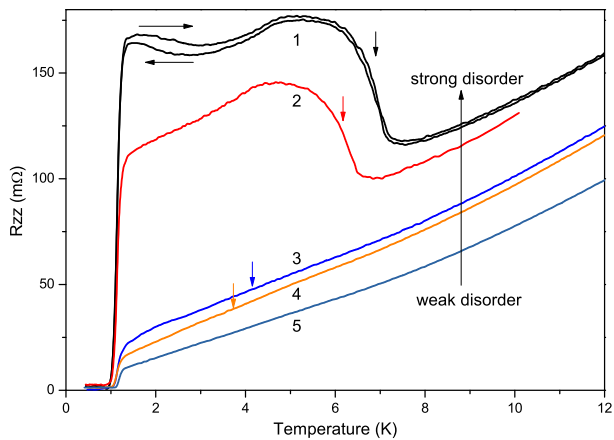}
\caption{Temperature dependences of interlayer resistance $R_{zz}$ for 5 different degrees of anion ordering. Arrows
indicate SDW onset as seen in $R_{xx}$.}
\label{fig3}
\end{figure}

The interlayer, $R_{zz}$, resistance behavior is quite different from the intralayer one, as one can see from Fig.~\ref{fig3}.
The most prominent difference consists in full superconducting transition for all the disorders, even those where SDW is
clearly observed in $R_{xx}$. In contrast to the two-step transition observed for $R_{xx}(T)$, e.g. \emph{d4}, characteristic
of a granular SC, the superconducting transition in $R_{zz}(T)$ is smooth. For strong disorders, the critical temperature
$T_c\approx1$\,K is only slightly reduced compared to the ordered sample $T_c\approx1.3$\,K\cite{Lebed08} and transition
becomes wider. There are evidences \cite{Joo04,Joo05,Yonezawa12} for nodal d-wave pairing in ClO4, therefore, $T_c$
suppression is expected with increasing disorder. Note, that disorder changes $R_{zz}$ almost by a factor of 3
(compare the \emph{d1} and \emph{d5} curves in Fig.~3 at $T=10$\,K).

The second difference is that the metallic behavior is preserved for $R_{zz}$ even for those disorders, where $R_{xx}$ is
clearly insulating. The abrupt increase of $R_{zz}$ resistance is observed for disorders \emph{d1,d2} at the same temperature,
where the the SDW onset is observed in $R_{xx}$. However, the increase is absent for disorders \emph{d3,d4} (cf.
Fig.~\ref{fig3}). Such a behavior might be expected due to the cross-section squeezing of the current paths as soon as
insulating background sets in.

The most remarkable result here is the hysteresis between the cooling and heating curves (cf. \emph{d1} curve on
Fig.~\ref{fig3}). Resistance is somewhat larger when the sample is heated from low temperatures, and the difference between
the cooling and heating curves vanishes only at $T_{SDW}$.  The hysteresis in $R_{zz}(T)$ is a stationary effect and is not
altered by either temperature sweep rate or the annealing time at $T<T_{SDW}$. Anion ordering is not expected to change at
temperatures much less than the gap. Indeed, annealing even at 10\,K has virtually no effect on the low-temperature
resistivity\cite{Schwenk84}. Due to strong disorder it is possible that some parts of the sample have lower local SDW transition
temperature and thus become insulating at lower temperatures than the majority of a sample bulk.  These structural
inhomogeneities, however, cannot account for the hysteresis.

Alternatively, the hysteresis might occur when phase transition between the two neighboring phases, namely between metal
and SDW is of the first order. In this case,  inclusions of the minority phase are present atop the majority phase even
below $T_{SDW}$. The amount of minority phase vanishes with decreasing temperature and is not restored until the phase
boundary is crossed again. The first order transition line was observed also by a number of probes in PF6
\cite{Vuletic02,Kornilov04,Lee05}, where it leads to quite similar temperature hysteresis behavior in $R_{zz}$\cite{Lee02} in
the SC/SDW coexistence region under pressure.

\begin{figure}[t]
\includegraphics[width=0.47\textwidth]{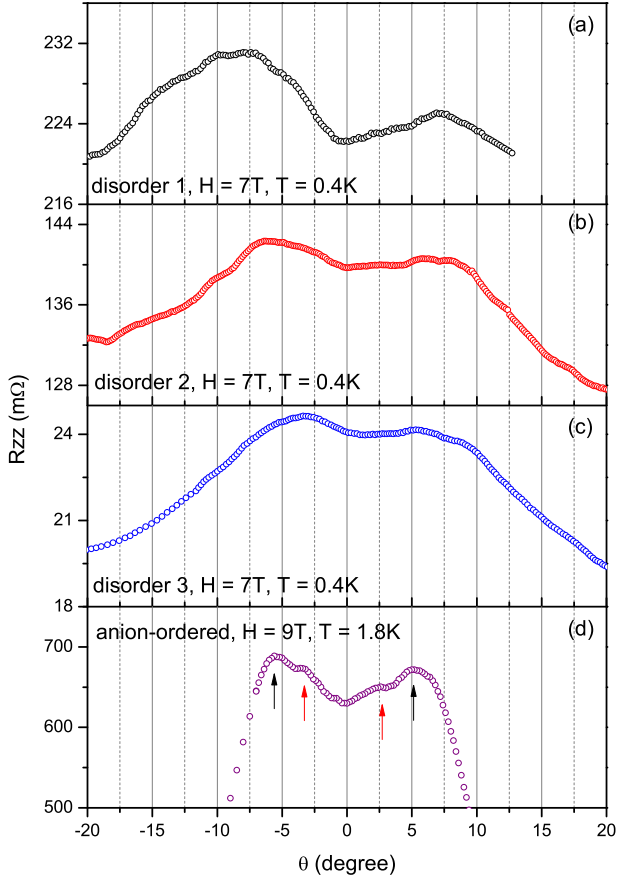}
\caption{ AMRO in the $\mathbf{a}-\mathbf{c}$ plane for disorders 1-3 and for the anion-ordered state (from top to
bottom). $\theta$ is the tilt angle of magnetic field with respect to the \textbf{a} axis in the $\mathbf{a}-\mathbf{c}$ plane
and is zero for  $H||\mathbf{a}$. Black and red arrows correspond to zeroth and first order oscillations. The disproportion of
the resistance in symmetric peaks is due to presence of small magnetic field $H||\mathbf{b}$, caused by slight misalignment of
the sample.}
\label{fig4}
\end{figure}

Deeper insight in the nature of metal and SDW coexistence can be obtained by  probing the influence of disorder on the energy spectrum.
Angular dependence of magnetoresistance is the major tool for obtaining information about the Fermi surface in quasi-one
dimensional metals. Figure~\ref{fig4} demonstrates $R_{zz}$  angular magnetoresistance oscillations (AMRO) for magnetic field
rotated in the $\mathbf{a}-\mathbf{c}$ plane. As one can see from Fig.~\ref{fig4}d, several pronounced peaks near
$H||\mathbf{a}$ are observed on the monotonic cosine background for a very weak anion disorder. This effect is well understood in
terms of Boltzmann transport and results from vanishing average electron velocity $\langle v_z\rangle$ on certain type of
orbits \cite{Danner94} (see Fig.~\ref{fig1}d). The average $\langle v_z\rangle$ has the same zeros as the Bessel function
$J_0(\frac{2t_b c}{\hbar v_f}\frac{B_x}{B_z})$, which corresponds to the orbits fitting the integer number of Brillouin zones
along $k_z$ (see cartoon on Fig.~\ref{fig1}d) \cite{Danner94}. One can see, that the angular distance between the peaks is
proportional to the $k_y$ bandwidth $2t_b$. In the presence of anion ordering the bandwidth is $t_b-V$ (see Fig.~\ref{fig1}c),
thus by measuring the AMRO effect in the $\mathbf{a}-\mathbf{c}$-plane we can directly probe the changes in local value of
$V$. The results are presented on Fig.~\ref{fig4} for a set of disorders, including those, where hysteresis in $R_{zz}(T)$
dependence is observed. The peaks are damped due to scattering time reduction at strong disorders, however zeroth-order peaks
are still present. We observe that with increasing disorder the peaks are moving away from each other, directly indicating
reduction of the band splitting $V$. Previous attempts\cite{Yoshino03} to observe the disorder dependence of $V$ used the AMRO
effect in $\mathbf{a}-\mathbf{b}$-plane. The latter effect is due to small enhancement of conductivity for certain directions
of magnetic field, nearly parallel to the velocities of the electrons in the inflexion points of FS \cite{Lebed08}. This small
enhancement is strongly damped by disorder thus making impossible observation of the  $\mathbf{a}-\mathbf{b}$ AMRO effect at
strong disorder.

X-ray diffraction measurements \cite{Pouget90} suggest that macroscopic anion-ordered grains are present even  for rapidly
cooled samples (i.e. at strong disorder). The SC ground state is certainly favorable inside the grains. Moreover, recent
theoretical studies \cite{Haddad11} also showed that d-wave SC can survive in ClO4 at strong disorder due to Josephson coupling
between SC inclusions, provided the distance between them is of the order of the coherence length. Therefore, it is important
to figure out the contribution of anion-ordered inclusions to the observed interlayer SC paths. The damped peaks in
the $\mathbf{a}-\mathbf{c}$ AMRO clearly indicate that the metallic/SC regions are disordered, and no additional peaks are
observed at the angular position of $\theta=\pm5.7^\circ$, corresponding to the anion-ordered state (see Fig.~\ref{fig4}). The
latter implies that either their overall contribution to conductivity is small or their size is comparable to the mean free
path. In any case, the interlayer SC paths are weakly influenced by the presence of anion-ordered inclusions.

\begin{figure}[t]
\begin{center}
\includegraphics[width=0.47\textwidth]{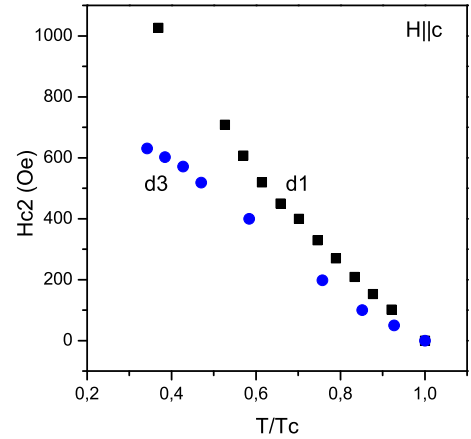}
\caption{Temperature dependence of critical field for disorders 1 and 3. The critical field was estimated as midpoint
($50\%$) of the resistive transition.}
\label{fig5}
\end{center}
\end{figure}

Another probe for the two-phase SC/SDW state is its superconducting critical field. Figure~\ref{fig5} shows the
$H_{c2}||\mathbf{c}$ temperature dependences for two disorders, \emph{d1} and \emph{d3} in the coexistence region. For stronger
disorders the $H_{c2}(T)$ curve shows a remarkable upturn, and the critical field becomes significantly enhanced
compared to disorder \emph{d3}. The positive curvature is absent for disorder \emph{d3}, as well as for the ordered
sample\cite{Lebed08}. This indicates, that the $H_{c2}$ upturn is a specific property of stronger disorders, where the presence
of SDW stronger affects the SC phase.

In the Ginzburg-Landau theory the slope $dH_{c2}/dT$ is related to the coherence lengths $\xi_i\xi_j$ in the plane normal to
the magnetic field direction.
 For a granular superconductor the increase of slope is possible provided the size of grains is smaller than $\xi$, however neither $R_{zz}(T)$ nor AMRO measurements demonstrate the significant contribution of the anion-ordered grains to the interlayer metal/SC paths. For homogeneous superconductor, the disorder induced variation of $V$ can change the effective mass and thus affect $\xi$. Indeed, the increase of the slope was observed for weak disorders \cite{Heidmann84}, but it was much weaker compared to that for disorder \emph{d1}.
On the other hand, the coexistence of SC and SDW phases is expected to affect SC
 either in real (e.g. when the size  varies due to  self-consistent formation of
spatially inhomogeneous state on the scale of $d\sim\xi$), or  in momentum space (e.g. when the effective mass  varies
due to FS reconstruction by nesting). Thus, critical field enhancement provides the solid evidence of the SC/SDW coexistence.

The strong enhancement of $H_{c2}||\mathbf{c^*}$ was observed in (TMTSF)$_2$PF$_6$\cite{Lee02} and was widely discussed
theoretically\cite{Zhang06,Grigoriev08,Grigoriev09}.  Some theories\cite{Zhang06,Grigoriev08} predict such an enhancement due
to the coupling of SDW and triplet SC order parameters. To our knowledge, no indication of triplet SC was ever observed in
ClO4, and for anion-ordered ClO4 recent angular dependent specific heat measurements in magnetic field are in support of the
nodal d-wave pairing\cite{Yonezawa12}. Unless the character of pairing is changed by the onset of SDW state
\cite{Grigoriev08}, the above scenarios \cite{Zhang06,Grigoriev08} are inapplicable to ClO4.

Another approach\cite{Grigoriev09} suggests that SDW order parameter is non-uniform with domain walls separating neighboring
regions with different order parameter. The domain walls can become superconducting, and critical field is enhanced due to
their small cross-section.  This approach also implies spatial anisotropy of the SC phase, since domain walls are normal to
the $\mathbf{a}$ axis. The latter is supported by recent transport anisotropy measurements in PF6\cite{Kang10}. Resistivity anisotropy
and critical field enhancement observed in this paper in ClO4 are also consistent with the domain walls scenario, though
additional measurements are required to prove it.

In summary,  using transport measurements and by varying anion disorder in (TMTSF)$_2$ClO$_4$ we probed the inner structure of
its complex state where the SDW phase coexists with  either SC phase (for $T<T_c$)  or metallic phase (for $T>T_c$).  The
hysteresis in temperature behavior of interlayer resistance points to the first-order character of metal-SDW phase boundary. We
find that the metallic/SC phase formed upon crossing this boundary is strongly spatially anisotropic. For strong disorders,  it
survives in a form of regions
 elongated at least in the interlayer ($\mathbf{c}$) direction, whereas along the $\mathbf{a}$ axis SDW insulating behavior is preserved. The
 critical field is enhanced in such a regions compared to the homogeneous SC state, which suggests that their cross-section
 decreases as SDW is stabilized by increasing disorder. At the same time, AMRO measurement in fields larger than $H_{c2}$ show
 that in the metallic/SC regions ClO$_4$ anions are predominantly disordered and band splitting $V$ is lower compared to the
 ordered state. Moreover, we directly observe that the latter is progressively reduced as disorder increases and SDW onset is
 shifted to higher temperatures.
The above results demonstrate that SC/SDW coexistence in ClO4 is driven by the FS splitting $V$ (similar to imperfect nesting caused by $t_b^\prime$ in PF6), rather than by structural inhomogeneities produced at the anion-ordering transition. This unifies the picture of SC/SDW
coexistence in ClO4 and PF6 or AsF6. Thus, ClO4 represents a peculiar example, where competition between magnetic insulating
and superconducting phases can be studied by continuous tuning of the degree of anion disorder.

Authors would like to thank P.D. Grigoriev and M.V. Kartsovnik for  discussion of the preliminary results. The work was supported by RFBR, Programs of the Russian Academy of Sciences, by  Russian Ministry for Education and Science (grant No 8375), and using research equipment of the Shared Facilities Center at LPI.

\end{document}